\documentclass[conference]{IEEEtran}
\ifCLASSINFOpdf
\else
\fi
\usepackage{svg}
\usepackage{amsmath}
\usepackage{graphicx}
\begin{document}

%
\title{Identification and Correction of False Data Injection Attacks against AC State Estimation using Deep Learning 
}

\author{\IEEEauthorblockN{Fayha ALmutairy}
\IEEEauthorblockA{Vermont Complex Systems Center\\ University of Vermont\\
Fayha.almutairy@uvm.edu\\}
\and
\IEEEauthorblockN{Reem Shadid}
\IEEEauthorblockA{ Applied Science Private University \\
re\_shadid@asu.edu.jo}
\and
\IEEEauthorblockN{Safwan Wshah}
\IEEEauthorblockA{Vermont Complex Systems Center\\ University of Vermont\\Safwan.Wshah@uvm.edu}}


%


\maketitle

\section{Abstract}
New advances in technology have greatly improved the monitoring and controlling of power networks, but these advances leave the system open to cyber attacks. One common attack is known as a False Data Injection Attacks (FDIAs), which poses serious threats to the operation and control of power grids. Hence, recent literature has proposed various detection and identification methods for FDIAs, but few studies have focused on a solution that would prevent such attacks from occurring. However, great strides have been made using deep learning to detect attacks. Inspired by these advancements, we have developed a new methodology for not only identifying AC FDIAs but, more importantly, for correction as well. Our methodology utilizes a Long-Short Term Memory Denoising Autoencoder (LSTM-DAE) to correct attacked-estimated states based on the attacked measurements. The method was evaluated using the IEEE 30 system, and the experiments demonstrated that the proposed method was successfully able to identify the corrupted states and correct them with high accuracy.

\section{Introduction}
 
Power systems and power grid networks are critical components in our modern-day electrical infrastructure \cite{Kush20018}. As such, ensuring both the physical and cyber security of these systems is of utmost importance. The use of information technology in power systems offers various benefits such as improved monitoring and controlling capability as well as allowed for the integration of renewable energy sources. But with these advantages come new risks. In recent years, cyber threats have been recognized as the most significant and dangerous threats to our power systems as a result of the increased dependency on information technology \cite{Gaoqi2013}. 
 
State estimation is an important power system application used to estimate the state of power transmission networks through a set of sensor measurements and network topology information \cite{Bobba2010}. The state estimator collects data about sensor measurements from the Supervisory Control and Data Acquisition (SCADA) system, including information about bus voltages, active and reactive power injection at each bus, and power flow in branches \cite{Ayad2018}.  It also removes all noise and error \cite{Kush20018}, as well as estimates the state of the power network\cite{Bobba2010}. These estimations are then used by network operators to ensure that the power grid is running in the desired states, allowing them to take corrective control actions if needed, and plan for any emergencies \cite{Bobba2010} \cite{Qingyu2014}. 
 
One of the most serious types of attacks on state estimation is a new one: False Data Injection Attacks (FDIAs). These attacks target the state estimation of the power grid by tampering with the data that are transmitted through communication lines \cite{Ashrafuzzaman2018} \cite{Nawaz2018}. FDIAs occur when an attacker introduces incorrect data or measurements into the system, which will reduce the control the system's operator has over the system, thus altering the state of the system \cite{Ayad2018}. It poses serious threats to the operations and control of power systems as it can evade the existing bad data detection (BDD) systems \cite{Sagnik2017}. In addition, FDIAs can affect deregulated energy markets and result in severe economic loses \cite{Kush20018}.

To address FDIAs, protection-based strategies, and detection-based strategies for defending the power systems have been proposed in the literature \cite{Ayad2018}. Multiple detection-based methods have been proposed. The most advanced methods use deep learning approaches such as Convolutional Deep Belief Network (CDBN) in \cite{Youbiao2017}, Recurrent Neural Network (RNN) in \cite{Ayad2018}, Deep Neural Network (DNN) in \cite{Ashrafuzzaman2018}, and Convolutional Neural Network (CNN) and Long-Short Term Memory (LSTM) in \cite{Xiangyu2018}\cite{Sagnik2017} to address the problem.  While substantial progress has been made in FDIAs detection, the complexity of the problem continues to present challenges to those in the power systems community. \cite{Kush20018} uses a statistical approach to address FDIAs, correcting measurements before feeding them to the state estimator.  One limitation of this method is the challenge of being scalable as it involves a complicated threshold scheme that would need to be applied to each scenario to detect the attack. Nevertheless, the scarcity of studies that have attempted to go a step further beyond attack detection and correct the errors resulting from FDIAs makes this field a promising one for future research \cite{Yanpeng2018}. 

Inspired by the significant success of deep learning (DL) methods in detecting attacks, in this paper, we present a new methodology for AC FDIAs identification and correction using DL to both identify and correct attacked measurements. With this method, we introduce the LSTM Denoising Autoencoder (LSTM-DAE) to reconstruct the signal such that it detects the anomaly and can then return it to its normal state. The new correction method was evaluated using the IEEE 30  system, and the experiments showed that the proposed method was successfully able to identify the corrupted states and correct them with high accuracy.

The remainder of this paper is organized as follows: Section II provides an overview of the state estimation process used in control centers and the false data injection attacks. Section III introduces the methodology of attack identification and correction using the LSTM-DAE. Section IV presents the case study, while Section V shows the simulation results of the case study with a brief discussion of the efficiency of the method used. Conclusions and suggestions for future work are presented in Section VI.

\section{Brief summary of State Estimation and FDIAs}

\subsection{State Estimation}
State estimation is a power system application used to estimate the state of power transmission networks through a set of sensor measurements and network topology information \cite{Bobba2010}. Network operators use the estimations of the current state to ensure that the power grid is running in the desired states, take corrective control actions, if needed, and plan for any emergencies \cite{Bobba2010} \cite{Qingyu2014}. 

There are two types of state estimation based on the power flow models of the network, namely DC and AC state estimations. A DC state estimation is a linear state estimation that is considered as a simplified version of the more complex AC or nonlinear state estimation. The major differences between DC and AC state estimations are: 1) in the DC state estimation, the solution is obtained in a closed form, while in the AC state estimation, it is obtained through iteration; 2) DC state estimation is based on active power flow analysis while the AC state estimation is based on reactive power flow analysis; and 3) in the DC state estimation, the only variables are the voltage phase angles, however, in the AC state estimation both the voltage phase angles and magnitudes are system variables \cite{Rahman2013}. 

For  DC state estimation, the relationship between the network's measurement and its state can be expressed by the following equation \cite{Bobba2010}:

\begin{equation}
z=Hx + e  
\end{equation}

where \(z=( z_1 , z_2, z_3, ...,z_m)^T\)is the sensor measurements, \( x = (x_1,x_2,...,x_n)^T\) is the true states of the system, ${H}$ is the Jacobian matrix where ${Hx}$ is a vector of m linear functions linking measurement to states, and \(e = (e_1,e_2,...,e_m)^T\) is the random errors in the measurement.

Since the DC model is a linear estimation of the network state, then the estimated system state ${x_{est}}$ can be expressed as:

\begin{equation}
{x_{est}}  =  (H^T WH)^{-1}H^TWz
\end{equation}

Where W is a diagonal matrix that consists of the measurement weights.
As for the AC state estimation, since the power flows are nonlinearly dependent, the relationship between the network’s measurement and its state can be expressed by the following equation \cite{Gabriela2012}:
\begin{equation}
z=h(x) + e                                            
\end{equation}

where ${h(x)}$ is a function vector that establishes the relationships between measured values and state variables.
Consequently, the estimated state variables are determined from the following optimization function:
\begin{equation}
min F(x) = (z - h(x))^T.W.(z-h(x)) 
\end{equation}

\subsection{FDIAs} 
False data injection attacks can be defined as those in which an attacker changes the sensor's measurements, which in turn changes the estimated value of the state variables without being flagged by the state estimator's bad measurement detection algorithms. FDIAs can be either random or targeted \cite{Bobba2010}. In random attacks, the attacker injects arbitrary errors into the estimates of state variables while in targeted attacks, specific errors are injected into the estimates of specific state variables.  A more significant categorization, however, is based on the extent of the knowledge that the attacker has about the network prior to launching the attack. As such, FDIAs can be either complete or incomplete.  

As the name suggests, with complete FDIAs, the attacker has complete knowledge of the power network's topology and transmission-line admittance values, allowing him or her to design a false data injection attack vector \cite{Rahman2012}. This type of attack is rare since the attacker is operating with limited resources and restricted physical access to the power grid. Incomplete attacks, the most common type, occurs when the attacker does not possess complete knowledge about the network's parameters and design \cite{Xuan2015}. For more information on complete and incomplete attacks see\cite{James2018}.

In this research, we deal with complete attacks, assuming that the attacker has the full knowledge required to carry out the attack. The attacker’s full knowledge makes detection of the attack more difficult.

\section{The Proposed LSTM-DAE Method}

 \begin{figure*}[t!] 
    \centering
    \includegraphics[width=\linewidth,keepaspectratio]{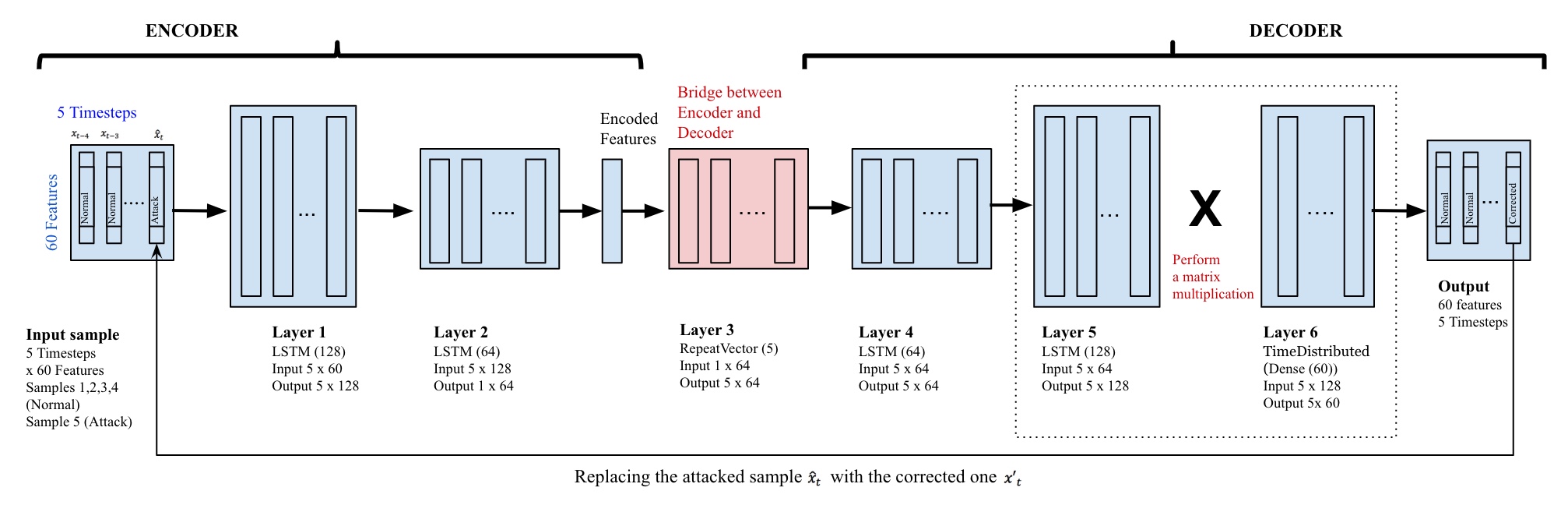}
    \caption{LSTM Denoising Autoencoder Architecture.}
    \label{fig:ARCH2}
\end{figure*}

 \begin{figure}[t] 
    \centering
    \includegraphics[width=80mm,keepaspectratio]{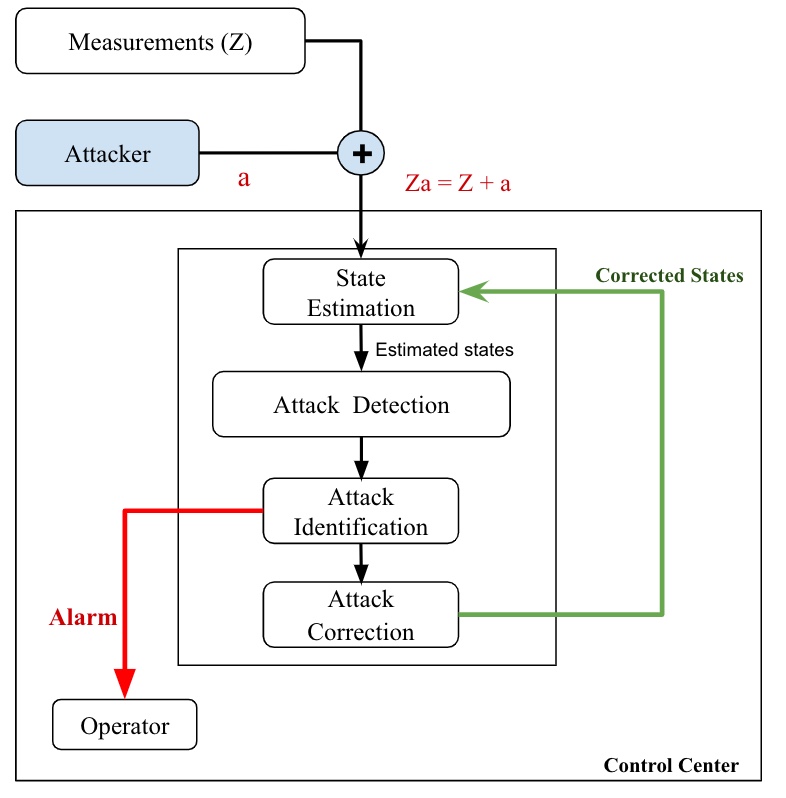}
    \caption{The Proposed Identification and Correction scheme for FDIAs.}
    \label{fig:ProposedM}
\end{figure}


Autoencoder, in general, is a neural network that is trained in order to learn a high-level representation of data by reducing data dimensionality. This is accomplished by having two main parts which are an encoder part \textbf{h = f(x)} and a decoder part \textbf{r = g(h)}. While vanilla Autoencoder is only useful to compress and decompress data without losing useful information \cite{Pierre2012}, Denoising Autoencoders (DAE) \cite{Vincent2008}, which is a special type of Autoencoder, can also learn to remove the noise from the corrupted input by recovering the original undistorted input. The DAE showed better representation more than the other Autoencoder types as the network learns a higher level of representation that is relatively robust to noise.

In this research, we aim to correct data affected by FDIAs. Inspired by the DAE and due to the sequential nature of the data, we are proposing a special type of RNN Autoencoder, namely, LSTM-DAE, to correct data after FDIA.  

To do so,  instead of feeding the original input ${x}$, we feed a noisy corrupted version of it $\hat{x}$, which is, in our case, the attacked states. Then, the loss function in that case is \textbf{L(x,g(f($\hat{x}$)))} as the reconstructed version of the original input ${x'}$ is computed from the corrupted input $\hat{x}$. 

LSTM, on the other hand, is a modified version of vanilla RNN, called gated RNN, which takes advantage of the sequential information to estimate the output \cite{Alex2018}, while having the ability to preserve long-term memory relations \cite{Hochreiter1997}. This is done by having three different types of gates, which are forget gate, input gate, and output gate. When they are combined together, the LSTM can be trained without worrying about losing information from the long-term memory or not capturing all the available information from the short-term memory.
Adding all this together, the proposed model consists of DAE stacking multiple layers of LSTM, where the input is a corrupted version of the data (Attack), and the output is the correct version of it (Normal).

The architecture is composed of three main parts, which are the encoder and the decoder, with a bridge between them. The first part consists of three parts, which are the input sample along with two LSTM layers. The input sample is a matrix whose dimensions are the timestep window \textit{w} and the number of states. The first LSTM layer of the encoder part consists of 128 units, and the Rectified Linear Unit (ReLU) is used as the activation function because it is the state-of-the-art activation function that can capture complex relations between the input and the output \cite{Vinod2010}. The second LSTM layer of the encoder part consists of 64 units with also a ReLU activation function. 

Following that, the bridge between the encoder and the decoder is represented by a Repeat Vector layer to reshape the output of the second LSTM layer in the encoder part so it can be fed into the first LSTM layer in the decoder part. This is achieved by repeating the output of the second LSTM layer in the encoder a number of times equal to the window size, which is 5 in this case.

For the decoder part, it starts with an LSTM layer with 128 units and a ReLU function as the activation function. This is followed by a Dense layer, which is a normal Neural Network, but it is wrapped in a Time Distributed layer to apply the Dense layer to every temporal slice of the input of the layer. Finally, the output of the Dense layer is the corrected states after training. The flow of the architecture can be seen in Figure \ref{fig:ARCH2}.

Our system works by correcting the samples online, such that the corrected sample is fed back to the states' queue in order to correct the next attacked state. Thus, the step window \textit{w} which represents the input of the model is divided into normal samples (\(x_{t-n},...,x_{t-3},x_{t-2},x_{t-1}\)) and an attacked sample ($\hat{x_t}$). To ensure that the model is immune to the noise coming from the corrected sample, Additive White Gaussian Noise (AWGN) is added to the states in the normal samples while training. This will guarantee that our system can handle the noise generated from the correction process. 

Our system assumes a perfect FDIAs detection; thus, our proposed model focuses on identification and correction. Our system identifies the attacked states by processing the corrected states from the proposed system. A direct comparison is applied between the corrected and attacked states after applying thresholds for theta and voltage to identify the attacked states. If any state value exceeds these thresholds, this indicates that state as an attacked state. The scheme of the process is illustrated in Figure \ref{fig:ProposedM}.

\section{Case Study}
In this research, we validated our results on the IEEE-30 bus system using real-world load data found on the Independent Electricity System Operator (IESO) website\cite{data}. Specifically, we trained our system on the data from 2002 to 2007 and tested it on the data from 2013 to 2015 to ensure that our system is not load dependent.

For FDIAs construction, we implemented the same methodology described in \cite{James2018}. To train and evaluate our system, ${w}$ timesteps samples were sequentially sampled from the normal dataset, and the last sample in the window has been attacked such that it is not detected by the state estimator. Thus, ${w-1}$ samples are normal ${x}$, and the last sample is the attacked state vector $\hat{x}$. Using this setup, 200,000 of ${w}$ samples have been constructed. We trained, evaluated, and tested our system with a ratio of 60/20/20, respectively.

For our experiments, we found that the best window size setting is 5. For the evaluation metric, we used the Root Mean Square Error (RMSE) to calculate the difference between the normal and the attacked values \cite{Chai2014}. Other metrics were also tested, such as Mean Absolute Error (MAE) and Mean Square Error (MSE), but RMSE gave the most easily interpretable results.

All the results and simulations were conducted on Nvidia K-80 GPU along with 16 GB of RAM. The LSTM-DAE was constructed using Keras, which is a top-level library built on top of TensorFlow for a computational speed boost.

 \begin{figure}[t] 
    \centering
    \includegraphics[width=\columnwidth,keepaspectratio]{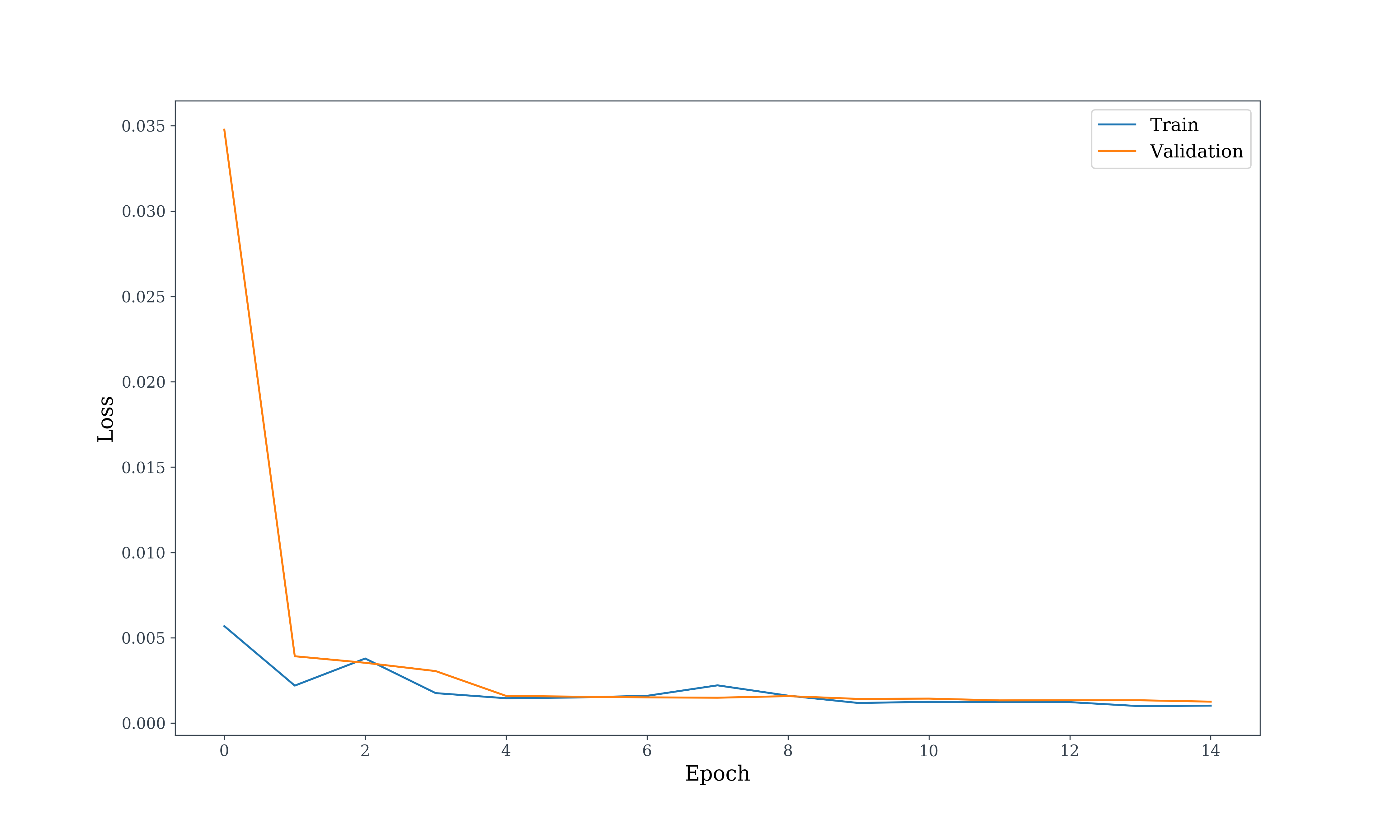}
    \caption{Model Loss (RMSE)}
    \label{fig:Results2}
\end{figure}
\begin{figure}[t] 
    \centering
    \includegraphics[width=\columnwidth,keepaspectratio]{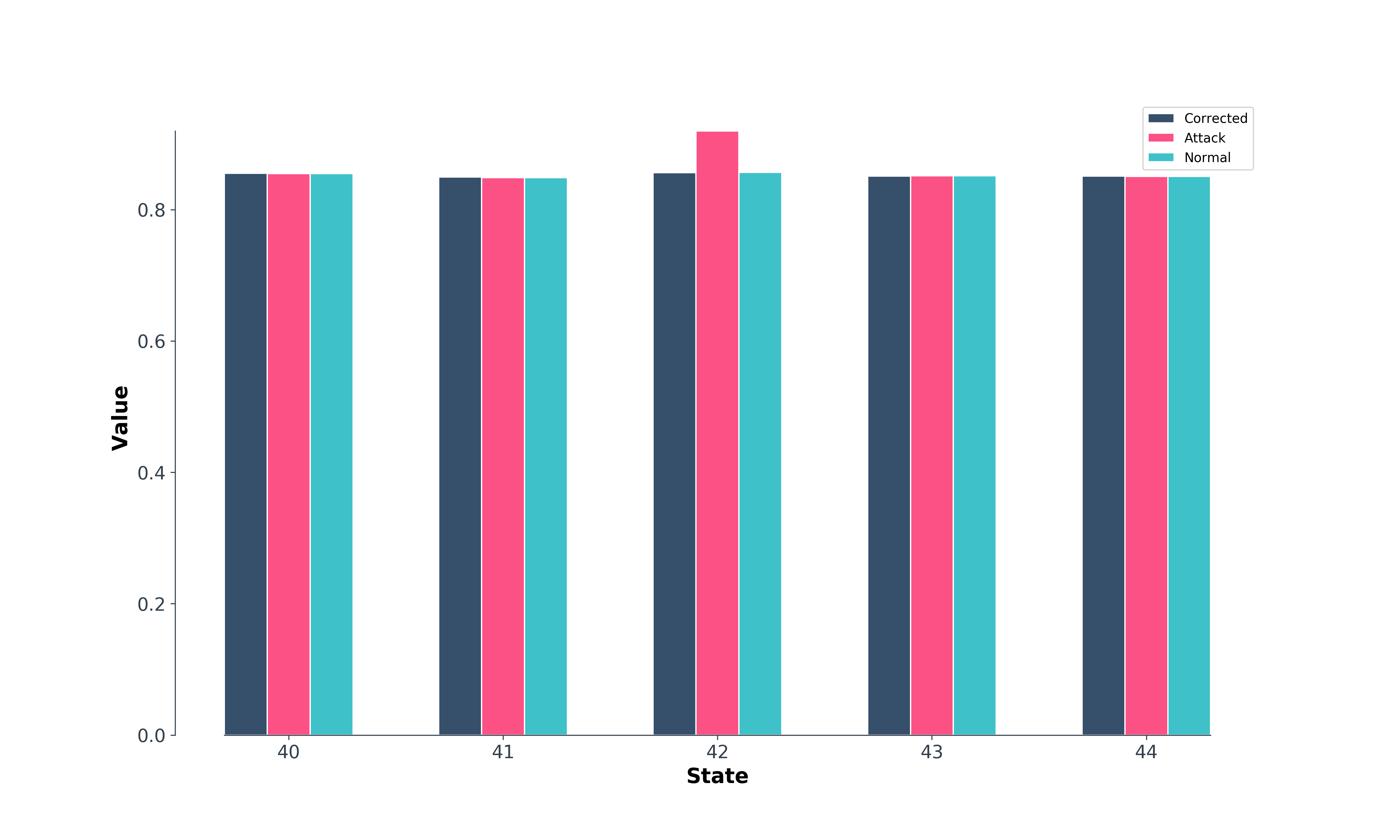}
    \caption{Reconstruction of a random sample for 5 random states}
    \label{fig:Reconstruction}
\end{figure}
\begin{figure}[t] 
    \centering
    \includegraphics[width=\columnwidth,keepaspectratio]{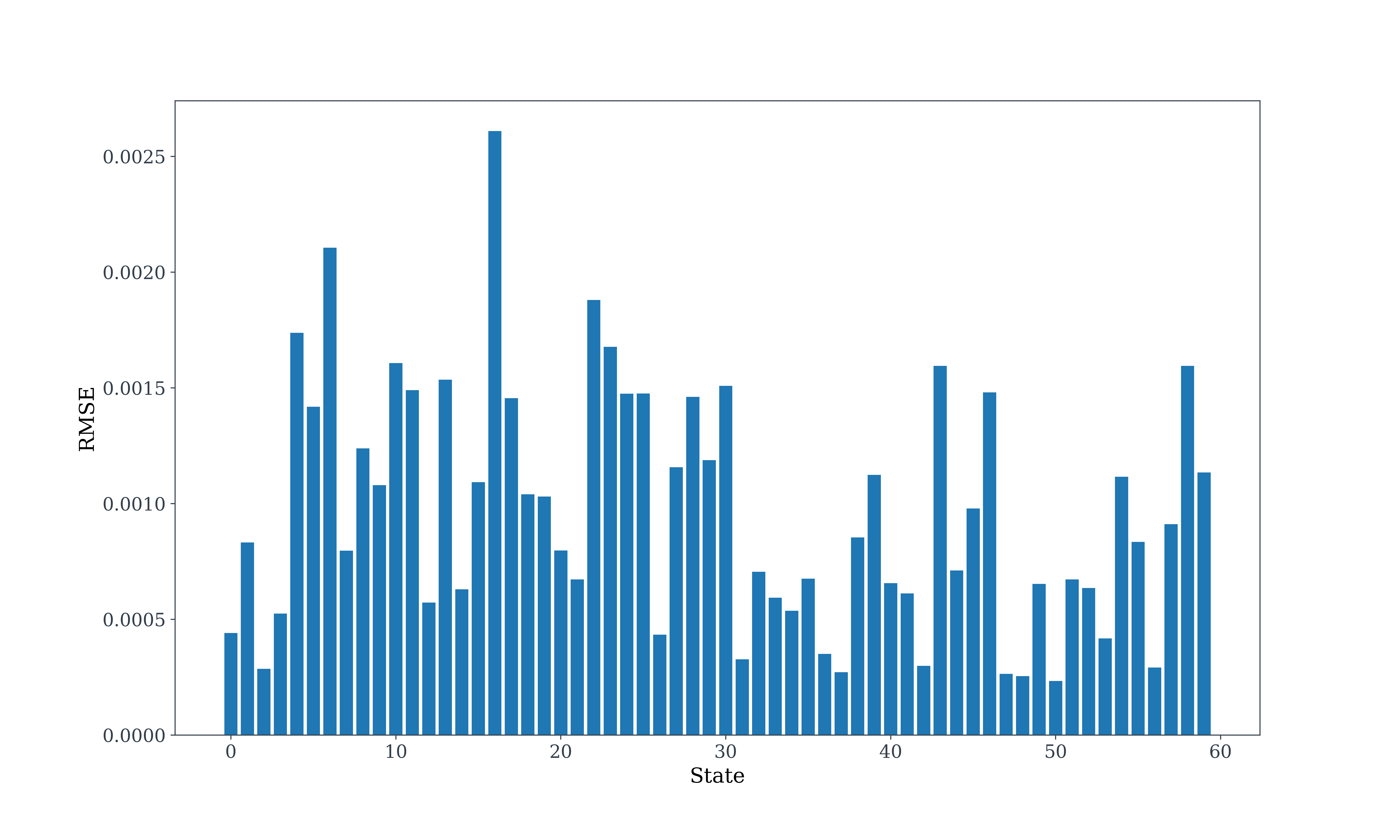}
    \caption{RMSE values for all states}
    \label{fig:RMSE_60}
\end{figure}

The results showed that our model is capable of capturing the temporal-spatial behavior of the data. As shown in Figure \ref{fig:Results2} , the loss of both the training and validation sets was monitored while training the model, and the results revealed that the RMSE was gradually decreasing until it reached less than 0.0013 on the validation data after 15 epochs. The RMSE reported on the testing set was 0.001 based on the best validation model.
Then, to validate that the model was capable of reconstructing any attacked state, a random sample was taken from the test set and the normal, attacked, and the corrected states’ values were measured. This is visualized in Figure \ref{fig:Reconstruction}, as the attacked state number 42 was successfully reconstructed with high accuracy.
Moreover, the RMSE values were calculated for all the 60 states to ensure once again that the model was capable of reconstructing any attacked state. The results in Figure \ref{fig:RMSE_60} show that this was achieved as the maximum RMSE was less than 0.0025, which is very low. 

\begin{figure}[t] 
    \centering
    \includegraphics[width=\columnwidth,keepaspectratio]{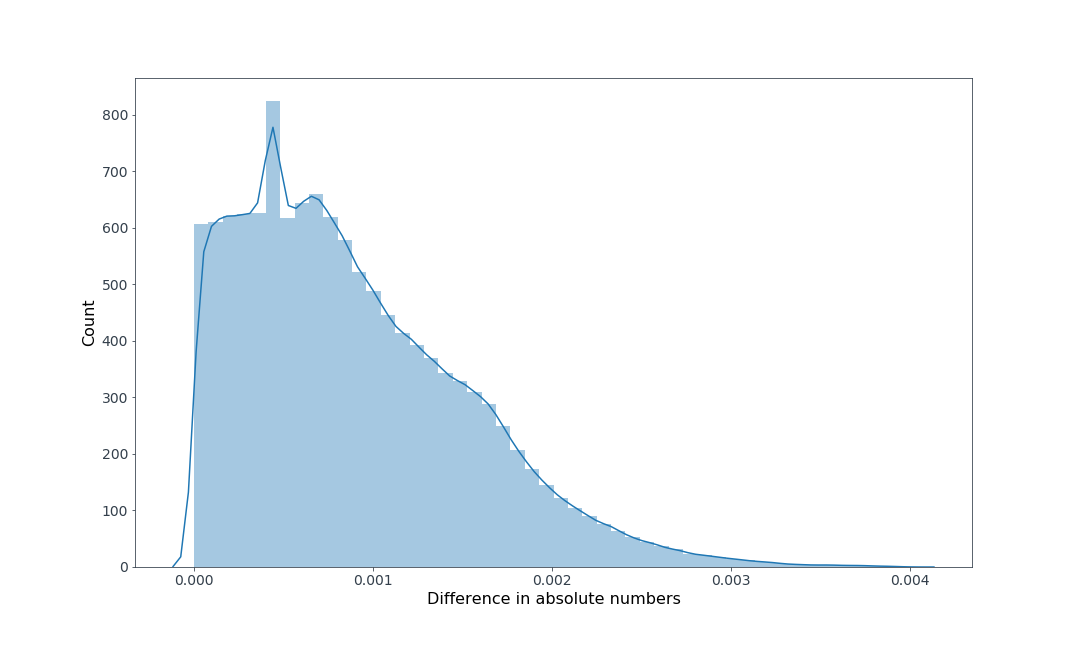}
    \caption{Histogram of the difference between actual and corrected statse.}
    \label{fig:Histogram}
\end{figure}

Finally, the difference between the actual states and the corrected one was calculated for all the states and timesteps in the test dataset. It was found that the maximum difference is less than 0.004, and most of the differences are between 0 and 0.002. In order to visualize this, a histogram graph was plotted in Figure \ref{fig:Histogram}. 

Surprisingly, our model is capable of almost reconstructing the attacked states to their normal values with a very low reconstruction error rate. Based on the reconstruction high accuracy, we were able to identify the source of the attack with a simple threshold mechanism,  which gave us 100 percent identification accuracy. We believe that our model has enough capacity to capture the behavior of the system in its steady-state given a real-world load on unseen load data. These very promising results have encouraged us to pursue research to model the state estimation itself using our approach.

\section{Conclusion}

In this paper, we proposed a new FDIAs identification and correction for AC state estimation. While some papers tackled the detection of FDIAs, little work has been done on identifying and correcting these attacks. The proposed method consists of an LSTM-DAE, which can capture the temporal-spatial dependencies. We validated our model on the IEEE 30-bus system using real-world load data. The results showed the model was able to identify the attacked states and correct them with high accuracy.


%
\IEEEpeerreviewmaketitle

\bibliographystyle{IEEEtran}
\bibliography{references}


\end{document}